\documentclass[epsfig, usenatbib]{mn2e}
\usepackage{epsfig}
\usepackage{natbib}
\usepackage{graphicx}
\usepackage{fixltx2e}
\usepackage{multirow}
\usepackage{amsmath}
\usepackage{amssymb}
\usepackage{url}
\title{The clustering of galaxies in the SDSS-III Baryon Oscillation Spectroscopic Survey: measuring structure growth using passive galaxies}

\author[Tojeiro et al.]{
 \parbox{\textwidth}{\LARGE
 Rita Tojeiro$^1$\thanks{E-mail: rita.tojeiro@port.ac.uk},  
 Will J. Percival$^1$,
 Jon Brinkmann$^2$, 
 Joel R.~Brownstein$^3$, 
 Daniel~J.~Eisenstein$^4$,
 Marc~Manera$^1$, 
 Claudia~Maraston$^1$,
 Cameron~K.~McBride$^4$,
 Demitri~Muna$^5$,
 Beth~Reid$^{6,7}$,
 Ashley~J.~Ross$^1$,
 Nicholas~P.~Ross$^6$,
 Lado~Samushia$^1$,
 Nikhil~Padmanabhan$^8$,
 Donald~P.~Schneider$^{9,10},$
 Ramin Skibba$^{11}$,
 Ariel~G.~S\'anchez$^{12}$,
 Molly~E.~C. Swanson$^4$,
 Daniel~Thomas$^1$,
 Jeremy~L.~Tinker$^5$,
 Licia~Verde$^{13}$,
 David~A.~Wake$^{8}$,
 Benjamin~A.~Weaver$^5$,
 Gong-Bo~Zhao$^{1,14}$
 }\vspace{3mm}\\
$^1$Institute of Cosmology and Gravitation, Dennis Sciama Building, University of Portsmouth, Burnaby Road, Portsmouth, PO1 3FX, UK \\
$^2$Apache Point Observatory, P.O. Box 59, Sunspot, NM 88349-0059, USA\\
$^3$Department of Physics and Astronomy, University of Utah, Salt Lake City, UT 84112, USA\\
$^4$Harvard-Smithsonian Center for Astrophysics, Cambridge, MA, USA\\
$^5$Center for Cosmology and Particle Physics, New York University, New York, NY 10003, USA\\
$^6$Lawrence Berkeley National Laboratory, 1 Cyclotron Road, Berkeley, CA 94720, USA\\
$^7$Hubble Fellow\\
$^{8}$Astronomy Department, Yale University, P.O. Box 208101, New Haven, CT 06520, USA \\
$^9$Department of Astronomy and Astrophysics, The Pennsylvania State University, University Park, PA 16802 \\
$^{10}$Institute for Gravitation and the Cosmos, The Pennsylvania State University, University Park, PA 16802 \\
$^{11}$Steward Observatory, University of Arizona, 933 N. Cherry Ave., Tucson, AZ 85721, USA \\
$^{12}$Max-Planck-Institut f\"{u}r extraterrestrische Physik, Postfach 1312, Giessenbachstr., 85741 Garching, Germany \\
$^{13}$ICREA \& ICC-UB University of Barcelona, Marti i Franques 1, 08028 Barcelona, Spain\\ 
$^{14}$National Astronomy Observatories, Chinese Academy of Science, Beijing, 100012, P.R.China\\
}

\def\gs{\mathrel{\raise1.16pt\hbox{$>$}\kern-7.0pt %
\lower3.06pt\hbox{{$\scriptstyle \sim$}}}}         %
\def\ls{\mathrel{\raise1.16pt\hbox{$<$}\kern-7.0pt %
\lower3.06pt\hbox{{$\scriptstyle \sim$}}}}         %

\newcommand{\vmatch}{$V_{\rm match}$ }

\newcommand{\mpcoh}{\,h^{-1}\,{\rm Mpc}}
\newcommand{\Hunit}{\,{\rm km}\,{\rm s}^{-1}\,{\rm Mpc}^{-1}}

\begin{document}

\maketitle

\begin{abstract}
{We explore the benefits of using a passively evolving population of galaxies to measure the evolution of the rate of structure growth between $z=0.25$ and $z=0.65$ by combining data from the SDSS-I/II and SDSS-III surveys. The large-scale linear bias of a population of dynamically passive galaxies, which we select from both surveys, is easily modelled. Knowing the bias evolution breaks degeneracies inherent to other methodologies, and decreases the uncertainty in measurements of the rate of structure growth and the normalization of the galaxy power-spectrum by up to a factor of two. 
If we translate our measurements into a constraint on $\sigma_8(z=0)$ assuming a concordance cosmological model and General Relativity (GR), we find that using a bias model improves our uncertainty by a factor of nearly 1.5. Our results are consistent with a flat $\Lambda$ Cold Dark Matter model and with GR.

%We measure $f\sigma_8(z)$, $b\sigma_8(z)$, $f(z)$ and $\sigma_8(z)$ with a precision of approximately 9-15\%, 2-3\%, 16-20\% and 6-9\% respectively, depending on redshift. }
}
\end{abstract}


\title{RSD measurements with passive galaxies}

\section{Introduction}  \label{sec:intro}

Current observational evidence points towards a Universe that is undergoing an accelerated expansion (see e.g. \citealt{KesslerEtAl09, AmanullahEtAl10, PercivalEtAl10, ReidEtAl10,BlakeEtAl11BAOb, BlakeEtAl11,BlakeEtAl11BAO,  ConleyEtAl11}). The physical reason behind such an acceleration remains poorly understood, and potential explanations range from a simple cosmological constant or vacuum density, to modified gravity models or an inhomogeneous Universe creating the illusion of an acceleration. Distinguishing between such physical explanations is a key goal of modern Cosmology. 

Redshift-space distortions (RSDs) are a key observational tool for understanding Dark Energy as they trace the matter velocity field via the peculiar velocities of galaxies. They allow a measurement of the growth rate of structure via an enhancement of the clustering power along the line of sight \citep{Kaiser87}. RSDs are powerful discriminants of different physical models for Dark Energy, as models that share the same expansion history often predict different growth rates of structure, $f$ (e.g. \citealt{LinderEtAl03}). 

Large-scale clustering measurements yield a direct measurement of $f\sigma_8$ and $b\sigma_8$, where $f$ is the logarithmic derivative of the linear growth factor $D(z)$ with the scale factor, $f\equiv d\log D(z)/d\log a$, $\sigma_8$ is the variance of the matter density field at a scale of $8$ $\mpcoh$, and $b$ is the large-scale linear galaxy bias. These results must be coupled with independent measurements of $b$ or $\sigma_8$ to yield an estimate of the growth rate, which often requires further assumptions: galaxy bias measurements are notoriously difficult, and measurements of $\sigma_8$ often need to be extrapolated in redshift.
Higher-order clustering measurements can also be used to break the
degeneracy between galaxy bias and cosmology (see e.g. \citealt{BernardeauEtAl02,ZhengEtAl07bias}) 
which has been investigated
with galaxy data (see e.g.  \citealt{PanEtAl05, GaztanagaEtAl2005, RossEtAl08, MarinEtAl11,McBrideEtAl11}).   Obtaining
precise constraints
from higher order moments is challenging, and this work serves as a
complement to
such investigations.  
	
In this paper we explore the gain if one knows the expected evolution of the bias for a sample of galaxies. For a passively evolving sample (i.e., no merging) the bias evolution can be computed using the formalism of \cite{Fry96} (see also \citealt{TegmarkEtAl98}). This formalism models a population of galaxies as being formed by a non-linear process at some time in the past, and subsequently evolving with the velocity flows set up by the matter density field. To first order (i.e. in the linear regime), a simple model for the evolution of bias can be constructed. This formalism assumes the continuity equation for the galaxy density field, which conserves the number of galaxies as a function of redshift, and imposes the need to select a dynamically passive sample. 

We obtain a passively evolving sample of galaxies via the method described in \cite{TojeiroEtAl12}, which provides weights and carefully matched galaxy samples spanning SDSS-I/II and III. 
Galaxies are weighted according to the volume in which they would be visible across the two surveys, and this matches the samples from an evolutionary point of view - SDSS-III galaxies seen through most of the SDSS-I/II volume are more likely to be their progenitors, and SDSS-I/II galaxies seen through most of the SDSS-III volume are more likely to be the evolutionary products of SDSS-III galaxies. One can then assess the consistency of this weighted sample with a dynamically passive model by computing the evolution of the number and luminosity densities - in a purely passive model, these should be constant with redshift. The most robust estimate of the merger rate in the weighted galaxy sample of \cite{TojeiroEtAl12} over $0.2<z<0.7$ yields a modest value of $2\%\pm1.5$\% Gyr$^{-1}$, establishing its suitability for our present study. 

When computing the large-scale clustering amplitude we weight each galaxy by its luminosity, and we construct samples at each redshift to have the same weighted luminosity density. The luminosity weighting gives a large-scale power estimator that is less sensitive to galaxies within the sample merging between any two redshifts: i.e. merging events between galaxies in the same halo do not affect the relative contribution of the halos within which they reside to the overall bias of the sample, provided total luminosity is conserved in such a merger. This is only strictly true in the case of no loss of light to the intra-cluster medium. Nonetheless, weighting by luminosity will almost always be better than any weighting scheme that depends on the number of objects in a halo - when two objects merge the relative contribution of a given halo to the overall clustering signal will be reduced by 1/2 if weighting by number. It follows that, provided that the overall loss of light is less than 50\% of the combined light of the merging system, we have an estimation of the bias evolution that is less sensitive to merging of galaxies within the sample, and to which the Fry model is more applicable.
The luminosity matching simply prevents selecting less luminous (and less biased) galaxies at different redshifts in case of merging. This would happen if one was to match samples on number density, for example.

It is these careful matching and weighting schemes that justify the use of the bias evolution of \cite{Fry96}. \cite{TojeiroEtAl12} further demonstrated that, assuming a $\Lambda$-Cold Dark Matter (${\Lambda}$CDM ) model and GR, the bias evolution of \cite{Fry96} provides a formally good fit to the data. Whereas in itself such a consistency is no proof of either the cosmological model or of the bias evolution model, it is a result that confirms our interpretation of the evolution of the galaxies within the broad context of a firmly motivated cosmological model. In this paper we assume the expansion history and matter power spectrum of a flat ${\Lambda}$CDM  universe, but we independently measure the growth rate of structure that gives the best fit to the data - which may be decoupled from the energy density and need not follow GR. The added constraint from the bias evolution allows us to break the degeneracy between galaxy bias, growth rate and $\sigma_8$.  

Finally we benefit from working on large scales (30-200 $\mpcoh$); the modelling of the matter power spectrum and RSDs on non-linear and quasi-linear scales is poorly understood and a further source of uncertainty (e.g. \citealt{ReidEtAl11}). In this first analysis we ignore most non-linear effects, accepting that future extensions of this work (with larger samples of galaxies and better statistical errors) will require a more sophisticated treatment of such effects.
Where required we assume a flat ${\Lambda}$CDM  cosmology with $\Omega_m = 0.25$, and $H_0 = 70\Hunit$ %km$^{-1}$Mpc$^{-1}$.

\section{Data}\label{sec:data}

The Baryon Oscillation Spectroscopic Survey (BOSS), as part of the Sloan Digital Sky Survey (SDSS) III \citep{EisensteinEtAl11}, increased the total SDSS-I/II imaging footprint to nearly $14,500$ sq. degrees; all of the imaging was re-processed as part of SDSS Data Release 8 \citep{AiharaEtAl11}.
In SDSS-I/II, Luminous Red Galaxies (LRGs) were selected for spectroscopic
follow-up according to the target algorithm described in
\citet{EisensteinEtAl01},  designed to follow a passive stellar population in colour and magnitude space. In SDSS-III,  the BOSS target selection extends the SDSS-I/II algorithm to target fainter
and bluer galaxies in order to achieve a galaxy number density of $3\times10^{-4}$ h$^3$ Mpc$^{-3}$ and increase the redshift range out to $z\approx0.7$. The spectroscopic footprint of the BOSS data used here covers $3275$ sq. degrees of sky, and corresponds to the upcoming Data Release 9, which will mark the first spectroscopic data release of BOSS. A set of comprehensive clustering analyses of this sample can be found in \cite{Aadvark, ReidEtAl12,SanchezEtAl12,ManeraEtAl12} and \cite{RossEtAl12}. The target selection algorithms for the LRGs and BOSS are described in detail in \cite{TojeiroEtAl12}. BOSS target selection consists of two separate algorithms - in this paper we use only the CMASS (for Constant MASS) sample, selected to be approximately stellar-mass limited, and targeting galaxies mainly with $z\gtrsim 0.43$ (Padmanabhan et al. in prep).

We split the data across four redshift slices: two slices of LRG galaxies centred at $z=0.3,0.4$ and two slices of the CMASS galaxies centred at $z=0.5,0.6$ ($\Delta z = 0.1$), with
$44136$, $30393$, $39780$ and $37883$ objects respectively. At each redshift we select the brightest galaxies until a fixed luminosity density is reached. This corresponds to roughly $95\%$ and $40\%$ of the LRGs and CMASS samples respectively.

\section{The model}\label{sec:model}

We describe the redshift-space galaxy correlation function $\xi(\mu,r)$ as in \cite{Hamilton92}:

\begin{equation}
\xi(\mu,r) = \xi_0(r)P_0(\mu) + \xi_2(r)P_2(\mu) + \xi_4(r)P_4(\mu)
\end{equation}
where $r$ is the comoving separation in $\mpcoh$ and $\mu$ is the cosine of the angle between a galaxy pair and the line of sight. $P_\ell$ are the Legendre polynomials with $P_0=1$, $P_2 = (3\mu^2-1)/2$ and $P_4=(35\mu^4- 30\mu^2+3)/8$.
$\xi_0$ is the monopole of the correlation function; the excess of finding a pair of galaxies at given distance $r$ averaged over pairs observed at all angles with respect to the line of sight. The quadropole, or $\ell=2$, contains the next order of information, by effectively comparing the power along and across lines of sight. Current  measurements of the octopole, or $\ell=4$, are too noisy to yield useful constraints and are not included in our model.
%
% and the following integrals are defined:
%
%\begin{equation}\label{eq:int1}
%\bar{\xi}(r) \equiv 3 r^{-3} \int_0^r \xi(r')r'^2dr'
%\end{equation}
%
%\begin{equation}\label{eq:int2}
%\bar{\bar{\xi}}(r) \equiv 5 r^{-5} \int_0^r \xi(r')r'^4dr'
%\end{equation}
%
%
We model the redshift evolution and the amplitude of the monopole and of the quadruple as \citep{Hamilton92}:

\begin{equation}\label{eq:A0}
\xi_0(r,z) = \left[ b^2(z) + \frac{2}{3}f(z)b(z) + \frac{1}{5} f^2(z) \right] \sigma^2_8(z)\xi_0^m(r)
\end{equation}
\begin{equation}\label{eq:A2}
\xi_2(r,z) = -\left[\frac{4}{3}f(z)b(z) + \frac{4}{7}f^2(z) \right] \sigma_8^2(z) \xi_2^m(r)
\end{equation}
with $\sigma_8(z) = \sigma_8(0)D(z)/D(0)$ where we set $D(0)=1$, and 

\begin{equation} \label{eq:fry96}
b(z) = [b(z_0) - 1] \frac{D(z_0)}{D(z)} + 1,
\end{equation}
where equation (\ref{eq:fry96}) follows the modelling of \cite{Fry96} for evolution of the large-scale linear bias. $\xi_{0,2}^m$ hold the information on the shape of the matter correlation function, and can be computed from $\xi^m(r)$ using a set of well-defined integrals (see \citealt{Hamilton92}). In this paper we use the $\xi^m_{0,2}(r)$ models of \cite{SamushiaEtAl12}, with $\Omega_m=0.25$.

We describe the three-equation system above with 4 parameters consisting of $b(z_0)$ and three nodes for $\sigma_8(z)$, which we model using a a quadratic polynomial. The nodes are at $z_{node}=0$, $0.3$ and $0.6$; we find that changing these nodes within this range does not affect our results significantly.

\section{The measurements}\label{sec:measurements}

We estimate the correlation function from the data, $\hat{\xi_\ell}(r)$,  by means of the \cite{LandySzalay93} estimator.
We use 130 bins in $r$, logarithmically spaced between $1$ and $200 \mpcoh$, and 200 linear bins in $\mu$, between 0 and 1. We use a random catalogue with the same angular mask as the data catalogue, and with an $n(z)$ matched to that of the data but with 10 times the number density. The non-trivial survey geometry imprints a non-uniform distribution of pairs in $\mu$ on the data. We correct for this effect as in \cite{SamushiaEtAl12}, by weighting each galaxy pair such that the weighted distribution of pairs in $\mu$ corresponds to that expected in the absence of a survey mask. We correct for angular and redshift completeness as in \cite{Aadvark}.

We weight each galaxy by its luminosity and \vmatch weight as described in \cite{TojeiroEtAl12}. The \vmatch weight preferentially selects galaxies seen across both surveys and more likely to belong to the coeval population of galaxies we wish to consider, and the luminosity weighting results in an estimate of the large-scale power that is less sensitive to merging within the sample - see Section \ref{sec:intro}. Together these weights ensure the bias model of Equation (\ref{eq:fry96}) is applicable to our sample. 

For each of the redshift slices we compute $\hat{\xi}_{0,2}(r)$, and use a simple 2-dimensional $\chi^2$ minimisation to find the best fitting scale-invariant amplitudes, $A_{0,2}(z)$, by writing $\hat{\xi}_{0,2}(r,z) = A_{0,2}(z)\xi^m_{0,2}(r)$. To ensure a stable inversion of the covariance matrix, and to increase our signal-to-noise in each bin, we re-bin $\hat{\xi}_{0,2}(2)$ to 11 bins between 30 and $200 \mpcoh$. Re-doing the analyses using scales between 50 and $200 \mpcoh$ significantly increases our overall errors, but does not change our conclusions.

We estimate the errors and their covariance by using mock simulations. We use the LasDamas mocks (McBride et al. in prep) to construct 80 independent realisations of $\hat{\xi}_{0,2}$ for the first two redshift slices (we sub-sample each mock in order to reproduce the $n(z)$ in each slice). For the last two redshift slices, we use 600 PTHalo mocks of \cite{ManeraEtAl12}, and follow the same procedure. We include the covariance between the multipoles in our fits. The CMASS mocks assume a slightly different ${\Lambda}$CDM  cosmology and are heavily subsampled to match the data $n(z)$; we scale their mean correlation function to match the data and apply the same factor squared to the full covariance matrix.

\section{Results} \label{sec:MCMC}

We adopt a Markov Chain Monte Carlo (MCMC) technique to sample the posterior distribution of the parameters in our model, given the data. We set uniform priors on our parameters as follows: $1 < b(z_0) < 3.5$ and $ 0 < \sigma_8(z_{node}) < 1.5$. The marginalised likelihood distributions of all our parameters have fallen to zero near these boundaries. 
We use a stationary proposal density function, with a shape similar to the marginalised likelihood distributions of each of our parameters, which we investigate with a set of preliminary chains. In each step of the chain we update one parameter at a time, randomly chosen and all with equality probability. Our final chains have an acceptance rate of $\approx 15\%$, and our results and $1\sigma$ intervals are robust to changes in the choice of the proposal and starting point; different choices for the proposal simply lead to lower acceptance rates. 
We adopt the mean value of each marginalised distribution as being the best-fit value for a given parameter, and we take $1\sigma$ errors from the standard deviation of the same distributions. 

\subsection{Passive model}
\begin{figure*}
\begin{center}
\includegraphics[angle=90, width=7.2in]{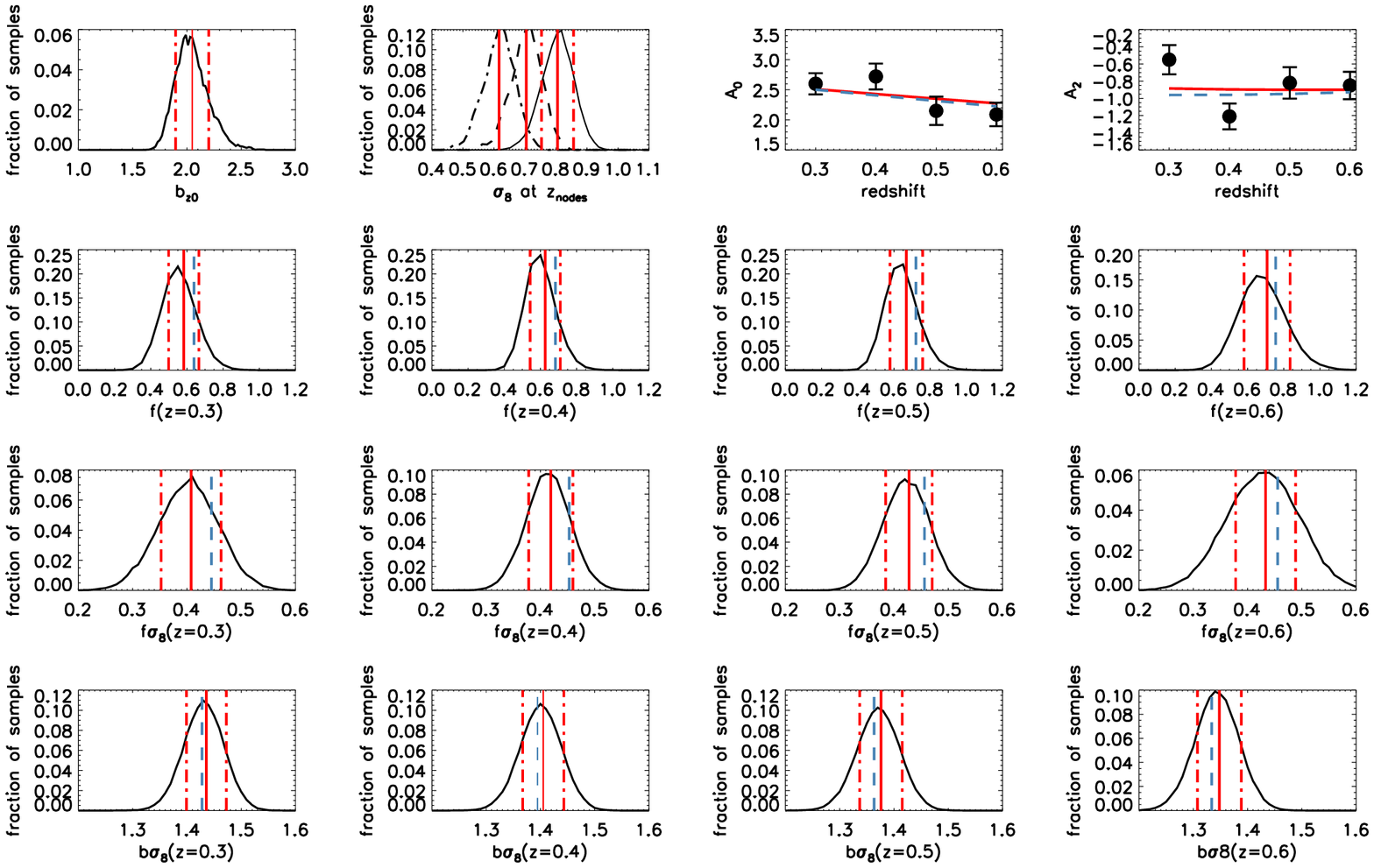}
\caption{Black curves in all panels show the marginalised likelihood distributions of our fitted and derived parameters. The fitted parameters are $b_{z_0}$ (first panel) and $\sigma_8(z_{nodes})$ (second panel, with $z_{node} = 0, 0.3$ and $0.6$ from right to left). The derived parameters are $f(z)$, $f(z)\sigma_8(z)$ and $b(z)\sigma_8(z)$. Vertical solid red lines show the best-fit values, and the vertical dot-dashed red lines the $1\sigma$ confidence intervals. Top right two panels show the measured value of $A_{0,2}(z)$ (black circles) and $1\sigma$ errors - the red line shows the best fit model. Dashed blue lines throughout show predictions from ${\Lambda}$CDM  and GR, using the best-fit values for the fitted parameters. GR is perfectly compatible with our measurements of the growth rate.} 
\label{fig:mcmc_constrained}
\end{center}
\end{figure*}

\begin{figure}
\begin{center}
\includegraphics[width=3.7in]{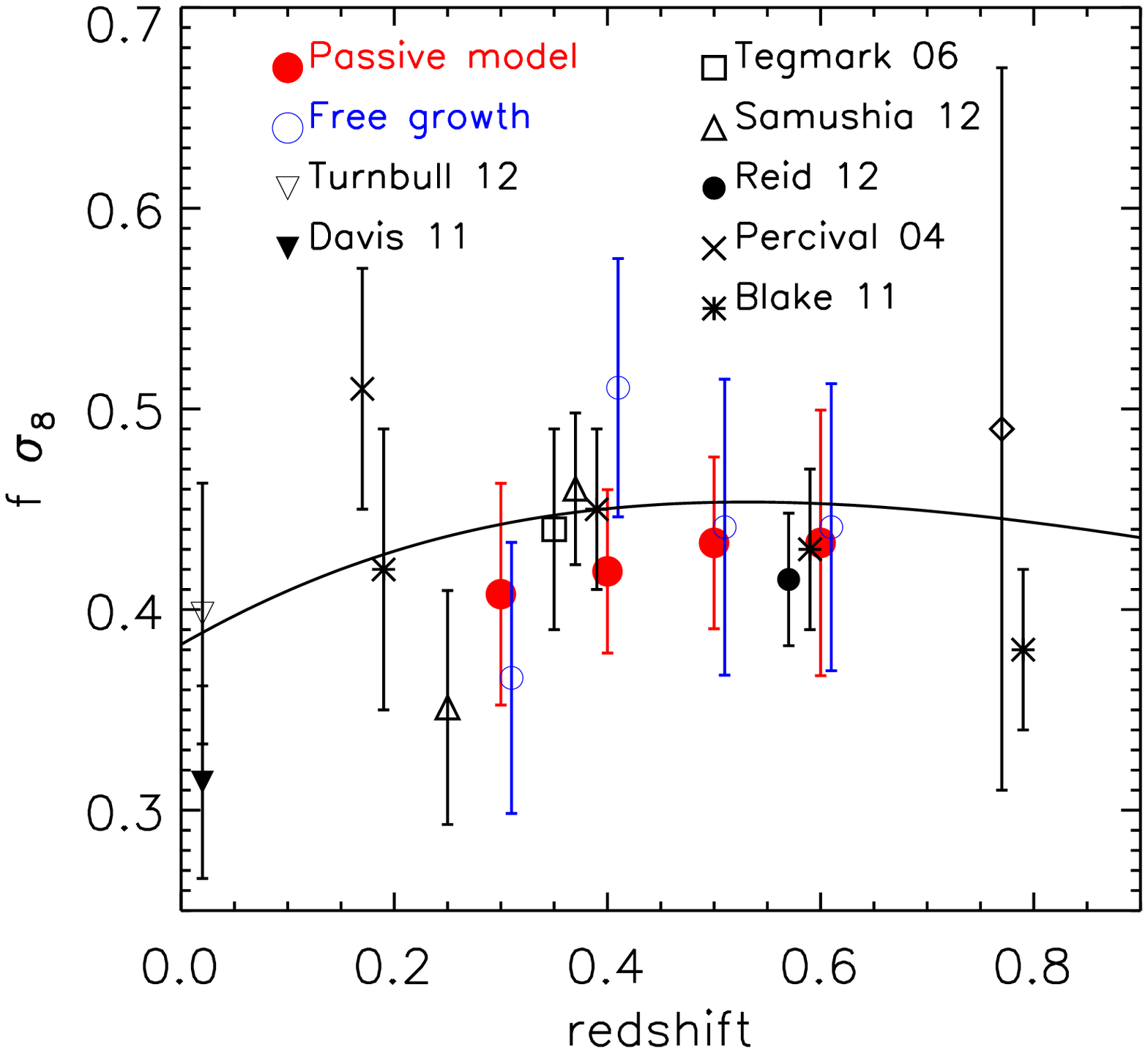}
\caption{Evolution of $f\sigma_8$ as a function of redshift for the passive model and free growth. The black data points are from: \protect\cite{BlakeEtAl11RSD},  \protect\cite{PercivalEtAl04RSD}, \protect\cite{TegmarkEtAl06} and \protect\cite{GuzzoEtAl08}; as collected by \protect\cite{SongEtAl09}. We also show measurements from \protect\cite{SamushiaEtAl12} and from \protect\cite{ReidEtAl12}. For completeness we also show the measurements of \protect\cite{DavisEtAl11} and \protect\cite{TurnbullEtAl12} from peculiar velocities at $z=0.02$, as compiled by \protect\cite{HudsonEtAl12}. The smooth solid line shows the prediction of ${\Lambda}$CDM  and GR, using a WMAP7 cosmology with $\sigma_8 (z=0) = 0.81$.} 
\label{fig:fs8_ev}
\end{center}
\end{figure}

Fig.~\ref{fig:mcmc_constrained} shows the marginalised likelihood distributions for the free parameters in our model: $b_{z_0}$ and $\sigma_8(z_{nodes})$ (first two panels), as well as for the derived parameters: $f(z)\sigma_8(z)$, $b(z)\sigma_8(z)$ and $f(z)$. We choose to present the distributions of the derived parameters at the centre of the redshift slices we use to measure the correlation function, but note that these are not independent. The correlation factor between adjacent measurements of $f(z)$ is high, between $0.84$ and $0.92$, but between the two furthest measurements, at $z=0.3$ and $z=0.6$, it is lower ($0.147$). The correlations of $f(z)\sigma_8(z)$ are similar. We show the best-fit values and $1\sigma$ confidence intervals in Table \ref{tab:summary}, under the header of passive model. The covariance matrix for our fitted parameters is given in Table \ref{tab:fitted} - this is the parameter set and covariance matrix that should be used for estimating likelihood surfaces. Fig.~\ref{fig:fs8_ev} shows in red our measurements of $f(z)\sigma_8(z)$ as a function of redshift, compared to measurements from the literature.

\begin{table*}
\begin{tabular}{|l|c|c|c|c|c|c|c|c|}
\hline \hline
\multirow{2}{*}{} & &\multicolumn{2}{|c|}{best-fit value} & \multicolumn{2}{|c|}{1$\sigma$ interval} & \multicolumn{2}{|c|}{\% error} \\  \cline{3-8} 
                             &  & passive model & free growth & passive model & free growth& passive model & free growtth  \\ \hline \hline
                             
\multirow{4}{*}{$f\sigma_8$} & $z=0.3$ & 0.407 &  0.366 & 0.055 & 0.067 & 13.55 & 18.3  \\
				            & $z=0.4$ & 0.419 & 0.511 & 0.041 & 0.064 & 9.71 & 12.5 \\
                       			     & $z=0.5$ & 0.427 & 0.447 & 0.043 & 0.073 & 10.01 & 16.3 \\ 
                       			     & $z=0.6$ & 0.433  & 0.441 & 0.067 & 0.071 & 15.27& 16.1 \\ \hline
\multirow{4}{*}{$b\sigma_8$} & $z=0.3$ & 1.436 & 1.438 & 0.037 & 0.062 &  2.56 & 4.31 \\
				            & $z=0.4$ & 1.405 & 1.417 & 0.037 & 0.068 & 2.61 & 4.80 \\
                       			     & $z=0.5$ & 1.376 & 1.321& 0.038 & 0.077 & 2.67 & 5.82 \\ 
                       			     & $z=0.6$ & 1.348 & 1.288 & 0.040 & 0.070 & 2.72 & 5.43 \\ \hline
\multirow{4}{*}{$f$} & $z=0.3$ & 0.582 & - & 0.094 &- & 16.1 & -  \\
			       & $z=0.4$ & 0.626 & - & 0.083 & - & 13.2& -\\
                            	     & $z=0.5$ & 0.668 & - & 0.090 & - & 13.5 & -\\ 
                 		     & $z=0.6$ & 0.708 & - & 0.127 & -  & 17.9 & -\\ \hline
$b$ & $z=0.3$ & 2.05 & - & 0.153 & - & 7.46 & - \\ \hline
\multirow{3}{*}{$\sigma_8$}  & $z=0$ & 0.804 & - & 0.051 & -  & 6.41 & - \\ 
 & $z=0.3$ & 0.704 & - & 0.049 & -  & 7.04 & - \\
 & $z=0.6$ & 0.617 & - & 0.050 & -  & 8.22 & - \\ \hline
\end{tabular}
\caption{Summary of the results in this letter. The passive model corresponds to the model described in Section \ref{sec:model}, using the bias evolution for passive galaxies. The free-growth model corresponds to the model described in Section \ref{sec:free_growth}.}
\label{tab:summary}
\end{table*}

\begin{table}
\begin{tabular}{|l|c|c|c|c|}
\hline
                  & $b_{z_0}$ & $\sigma_8(0)$& $\sigma_8(0.3)$& $\sigma_8(0.6)$ \\ \hline
 $b_{z_0}$ & 0.02335        &                  - &   - & - \\
 $\sigma_8(0)$ & -0.006917 &   0.002666 &    - &  - \\
 $\sigma_8(0.3)$ & -0.007086 &    0.002338 &    0.002459 &   - \\
 $\sigma_8(0.6)$ & -0.007000 &    0.002293 &    0.002482 &   0.002570 \\ \hline
 \end{tabular}
 \caption{Covariance matrix for the fitted parameters recovered from the MCMC chain described in Section \ref{sec:MCMC}.}
 \label{tab:fitted}
 \end{table}

\subsection{Free growth model}\label{sec:free_growth}

To place the results from the previous Section into context, we fit $f\sigma_8$ and $b\sigma_8$ independently in each of the redshift slices. We continue to use equations (\ref{eq:A0}) and (\ref{eq:A2}), but now drop the constraint on the bias evolution given by (\ref{eq:fry96}). We use an MCMC similar to the one described in Section \ref{sec:MCMC}, adapted to reflect the different parameters in this model, of which there are eight. The evolution of $f\sigma_8$ can be seen in the blue points of Fig.~\ref{fig:fs8_ev}, and we show the full set of results in Table \ref{tab:summary} under the header of free growth.
%When comparing like for like, there is an apparent gain in precision of up to a factor of two when using the passive model. Note that these points are now independent, if we assume that the measurements of $A_0(z)$ and $A_2(z)$ are also independent across the redshift slices (we compute the intra-multipole covariance using the mock catalogues). 
We see a loss in precision of up to a factor of two in the estimation of $f(z)\sigma_8(z)$ and $b(z)\sigma_8(z)$, when compared to the constraints obtained using the passive model. Note that the measurements quoted under free growth in Table~\ref{tab:summary} at each redshift are now independent. %We assess the constraining power of the two sets of results in the next Section.

\subsection{Constraining power}\label{sec:constraining}
\begin{figure}
\begin{center}
\includegraphics[angle=90, width=3.5in]{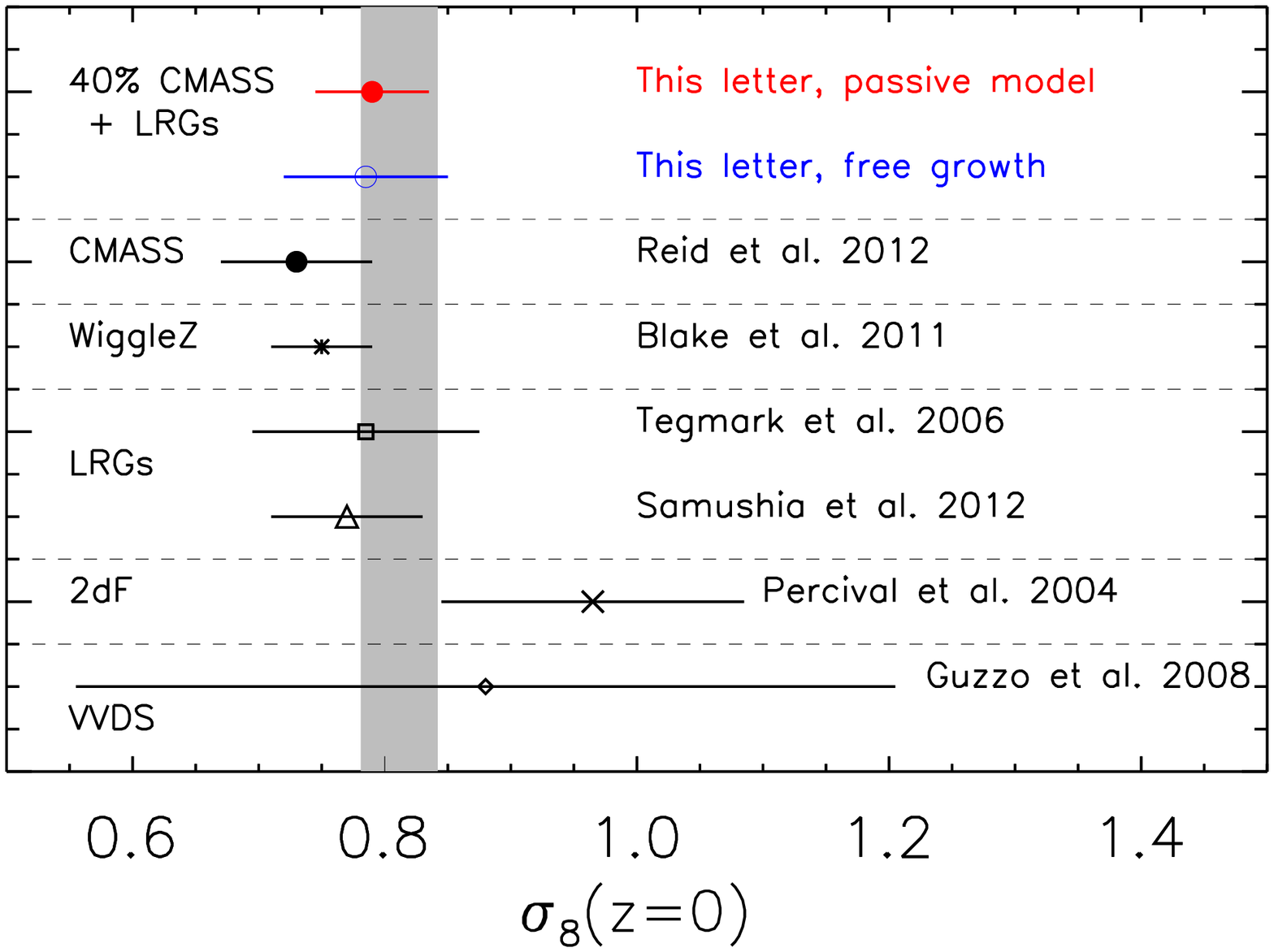}
\vspace{2mm}
\caption{Constraints on $\sigma_8(z=0)$ from the data points in Fig.~\ref{fig:fs8_ev}, assuming ${\Lambda}$CDM  and GR. The vertical shaded bar shows the constraints placed by the joint data
analysis in WMAP7 \protect\citep{KomatsuEtAl11}. The constraints from the passive model are approximately 1.5 times better than a free growth model, and competitive relative to  \protect\cite{ReidEtAl12} on the full CMASS sample. On the left we show the dataset used for each measurement.} 
\label{fig:s8_constraints}
\end{center}
\end{figure}

As it is difficult to judge the constraining power of correlated measurements, we undertake the following exercise. Assuming GR and ${\Lambda}$CDM , we assess how well $\sigma_8(z=0)$ can be constrained, using each set of points in Fig.~\ref{fig:fs8_ev}. When using literature data, we assume the likelihood surfaces to be gaussian, and in the case of multiple measurements we assume them independent. In the case of the measurements derived in this Letter, we use the best-fit $\sigma_8 (z_{nodes})$ values and their covariance. We show the resulting constraints in Fig.~\ref{fig:s8_constraints}. The constraints from the passive model are approximately 1.5 times better than a free growth model, and competitive when compared to state-of-the-art results of \protect\cite{ReidEtAl12} on the full CMASS sample, and \protect\cite{BlakeEtAl11RSD} with WiggleZ.

\section{Summary and conclusions}\label{sec:summary}

We demonstrate for the first time how using a passive sample of galaxies can enhance the accuracy of the measurement of the growth rate, via the added knowledge of the evolution of the large-scale galaxy bias. Our results are fully consistent with a flat $\Lambda$CDM model and GR. When compared to fitting $b\sigma_8$ and $f\sigma_8$ independently at each redshift, we find an increase in precision of up to a factor of two. If we translate our ${\Lambda}$CDM measurements into a constraint on $\sigma_8(0)$, assuming ${\Lambda}$CDM  and GR, we find that a passive model gives $\sigma_8(0)=0.79\pm0.045$ which is a nearly 1.5 times improvement on the results obtained using a free growth model, $\sigma_8(0) = 0.785\pm0.065$. Furthermore, these constraints are comparable with those obtained using the measurement of \cite{ReidEtAl12}, $\sigma_8(0) = 0.755^{+0.065}_{-0.060}$, whilst only using $\sim 40\%$ of the BOSS CMASS galaxies (but adding SDSS-I/II). This technique offers great potential, and it will deliver highly competitive results as BOSS gathers more data. 

A smaller statistical error in the measurements will require a more sophisticated modelling of non-linearities in the treatment of RSDs, as well as a potential extension of the bias evolution model to accommodate a sample of galaxies that will be increasingly less dynamically passive as we extend this work in luminosity and/or redshift.
The obvious caveat is that we need to provide a convincing case that a sample is well matched to passive evolution. For our sample this was provided by \cite{TojeiroEtAl12}. 

With the right dataset and modelling, it is straightforward to extend this technique to higher redshift, and map the growth of structure over a larger fraction of the age of the Universe.
%%%%%%%%%%%% Acknowledgments %%%%%%%%%%%%%%%%%%%
\section{Acknowledgments}

RT and WJP are thankful from support from the European Research Council. MECS was supported by the National Science
Foundation under Award No. AST-0901965. Funding for SDSS-III has been provided by the Alfred P. Sloan Foundation, the Participating Institutions, the National Science Foundation, and the U.S. Department of Energy. The SDSS-III web site is http://www.sdss3.org/.

SDSS-III is managed by the Astrophysical Research Consortium for the
Participating Institutions of the SDSS-III Collaboration including the
University of Arizona,
the Brazilian Participation Group,
Brookhaven National Laboratory,
University of Cambridge,
Carnegie Mellon University,
University of Florida,
the French Participation Group,
the German Participation Group,
Harvard University,
the Instituto de Astrofisica de Canarias,
the Michigan State/Notre Dame/JINA Participation Group,
Johns Hopkins University,
Lawrence Berkeley National Laboratory,
Max Planck Institute for Astrophysics,
Max Planck Institute for Extraterrestrial Physics,
New Mexico State University,
New York University,
Ohio State University,
Pennsylvania State University,
University of Portsmouth,
Princeton University,
the Spanish Participation Group,
University of Tokyo,
University of Utah,
Vanderbilt University,
University of Virginia,
University of Washington,
and Yale University.
\vspace{-3mm}
%%%%%%%%%%%%% Biblio %%%%%%%%%%%%%%%%%%%
\bibliographystyle{mn2e}
\bibliography{../PAPER/my_bibliography}

\end{document}